\documentclass[aps,pra,reprint,superscriptaddress,floatfix]{revtex4-2}
\usepackage{graphicx}
\usepackage{dcolumn}
\usepackage{bm}
\usepackage{amsmath}
\usepackage{overpic}
\usepackage{subfigure}
\usepackage{ragged2e}
\usepackage{amssymb}
\usepackage{comment}

\usepackage[colorlinks=true,citecolor=blue,urlcolor=blue]{hyperref}

\begin{document}
\graphicspath{{Figures/}}

\title{Reflection and Refraction at Nonlinear Temporal Boundaries in Synthetic Lattices}

\author{Chong-Xiao Chen}
\affiliation{Anhui Province Key Laboratory of Quantum Network, University of Science and Technology of China, Hefei 230026, China}
\affiliation{CAS Center For Excellence in Quantum Information and Quantum Physics, University of Science and Technology of China, Hefei 230026, China}

\author{Zheng-Wei Zhou}
\affiliation{Anhui Province Key Laboratory of Quantum Network, University of Science and Technology of China, Hefei 230026, China}
\affiliation{CAS Center For Excellence in Quantum Information and Quantum Physics, University of Science and Technology of China, Hefei 230026, China}
\affiliation{Hefei National Laboratory, University of Science and Technology of China, Hefei 230088, China}
\affiliation{Anhui Center for Fundamental Sciences in Theoretical Physics, University of Science and Technology of China, Hefei, 230026, China}

\author{Xi-Wang Luo}\email{luoxw@ustc.edu.cn}
\affiliation{Anhui Province Key Laboratory of Quantum Network, University of Science and Technology of China, Hefei 230026, China}
\affiliation{CAS Center For Excellence in Quantum Information and Quantum Physics, University of Science and Technology of China, Hefei 230026, China}
\affiliation{Hefei National Laboratory, University of Science and Technology of China, Hefei 230088, China}
\affiliation{Anhui Center for Fundamental Sciences in Theoretical Physics, University of Science and Technology of China, Hefei, 230026, China}

\date{\today}

\begin{abstract}
Temporal boundaries in time-modulated media provide a powerful route toward wave manipulation beyond conventional spatial boundaries. Here, we investigate nonlinear temporal boundaries generated by interaction quenches in a synthetic lattice with exactly solvable interacting dynamics. Unlike conventional temporal boundaries arising from abrupt changes of single-particle dispersion, the present system realizes a self-induced temporal medium in which the propagating wave packet dynamically determines its own effective dispersion and transport properties. By solving the nonlinear Schrödinger dynamics analytically, we show that the interaction generates an emergent wave-packet-dependent band structure and a state-dependent temporal refractive response while preserving fully controllable evolution. Based on this framework, we establish a nonlinear temporal-scattering picture and uncover phenomena including amplitude-dependent temporal reflection/refraction and nonlinear temporal birefringence.
Furthermore, we demonstrate that gradient-induced Bloch oscillations suppress wave-packet diffusion and enable coherent periodic transport with exact state reconstruction. Our results extend temporal reflection and refraction from dispersion-quenched linear systems to interaction-quenched nonlinear media and provide a tractable framework for nonlinear wave manipulation in synthetic lattices.

\end{abstract}

\maketitle
\section{Introduction}
Temporal boundaries in temporally modulated media have recently emerged as a powerful paradigm for wave manipulation beyond conventional spatial boundaries~\cite{akhmanov1969nonstationary,Kolner1994SpacetimeDuality,Mendonca2002TimeRefraction,morgenthaler2003velocity,Plansinis2015WhatTemporal}. Unlike spatial boundaries, which conserve energy while modifying momentum, temporal boundaries preserve momentum while allowing energy exchange with the external temporal drive, leading to unconventional scattering processes governed by temporal analogs of Snell's laws~\cite{Hannaford2024ReflectionRefraction,Mai2023FundamentalAsymmetries,Dong2024QuantumTime}.
This distinct conservation principle gives rise to a variety of remarkable phenomena, including frequency conversion, time reversal, and temporal reflection/refraction~\cite{Bacot2016TimeReversal,Kim2024TemporalRefraction,Xiao2014ReflectionTransmission,Lustig2021PhotonicTimecrystals,Zhou2020BroadbandFrequency,Long2023TimeReflection,Dong2024QuantumTime}. Experimental demonstrations in acoustic and water-wave systems~\cite{Bacot2016TimeReversal,Kim2024TemporalRefraction}, as well as in epsilon-near-zero materials such as indium tin oxide (ITO)~\cite{Zhou2020BroadbandFrequency}, have confirmed key signatures of temporal scattering. Nevertheless, realizing sharp temporal boundaries in conventional media remains challenging because finite material response times constrain both the modulation speed and modulation depth, thereby limiting the strength and controllability of temporal scattering effects~\cite{Lustig2021PhotonicTimecrystals,Zhou2020BroadbandFrequency,Long2023TimeReflection}.

In parallel, the development of synthetic dimensions has established highly controllable and scalable platforms for engineering wave dynamics~\cite{Celi2014SyntheticGaugea,Gadway2015AtomopticsApproacha,Mancini2015ObservationChiral,Ozawa2016SyntheticDimensionsa,Ozawa2019TopologicalQuantum,Yuan2018SyntheticDimensiona,Chen2024StronglyInteracting,Kanungo2022RealizingTopological}. Synthetic lattices constructed from internal atomic or photonic degrees of freedom provide precise control over band structures, transport properties, and interactions, enabling studies of a broad range of quantum and classical wave phenomena~\cite{Wang2023TestingUniversality,An2018CorrelatedDynamics,Xie2020TopologicalQuantum,Chen2025InteractiondrivenBreakdown,Zhang2025ObservationHigherorder,Xu2025ProbingBulk}. More recently, synthetic-dimensional platforms in photonic and ultracold-atom systems have provided an alternative route toward realizing sharp temporal boundaries, enabling experimental observations of temporal reflection and refraction through controlled engineering of single-particle dispersions~\cite{Long2023TimeReflection,Dong2024QuantumTime}.
In these studies, the temporal boundaries are generated by sudden changes in the single-particle band structure and therefore represent a fundamentally linear type of temporal boundary. Such systems have established the basic principles of temporal scattering in synthetic lattices and demonstrated excellent controllability compared with conventional temporally modulated media.

A natural next step is to ask whether temporal scattering can be extended beyond the linear regime by creating temporal boundaries through interaction quenches. In contrast to linear temporal boundaries, where only the single-particle dispersion changes across the boundary, an interaction-induced temporal boundary introduces intrinsically nonlinear and state-dependent dynamics. Since the nonlinear propagation properties depend on the wave itself~\cite{Boyd2023,AgrawalNonlinear}, a sudden change of the interaction strength can, in principle, generate temporal scattering processes whose characteristics depend on the intensity and profile of the incident wave packet. Such nonlinear temporal boundaries could therefore support fundamentally new forms of temporal reflection and refraction that have no counterpart in linear systems.

Despite this intriguing possibility, nonlinear temporal boundaries remain largely unexplored. Initial considerations have been discussed in momentum-space lattices (see Supplementary Information of Ref.~\cite{Dong2024QuantumTime}) and in the context of ``Bose-fireworks'' dynamics in trapped Bose-Einstein condensates~\cite{Clark2017CollectiveEmission,Feng2019CorrelationsHighharmonic,Chen2020ManybodyEcho}. However, interaction quenches in nonlinear systems typically generate modulational instabilities, mode mixing, and complex nonequilibrium dynamics. As a result, the ensuing evolution often cannot be described in terms of simple scattering processes, making it difficult to identify clear analogs of temporal reflection and refraction. Developing a controllable framework for nonlinear temporal scattering that retains both strong state dependence and analytical transparency therefore remains an outstanding challenge.

Here, we investigate temporal scattering induced by a sudden interaction quench in a synthetic lattice with a nonlocal interaction. A key feature of this interaction is its locality in Bloch momentum space, which renders the nonlinear Schr\"odinger equation exactly solvable. We derive an exact analytical solution for the evolution of an arbitrary initial wave packet, including in the presence of a gradient potential that drives Bloch oscillations. This solution reveals an emergent effective band structure whose properties depend self-consistently on the profile of the wave packet, allowing the nonlinear dynamics to be interpreted in terms of wave-packet-dependent band transport. Building on this framework, we demonstrate amplitude-dependent temporal reflection and refraction, nonlinear temporal birefringence, and coherent transport with diffusion suppression enabled by Bloch oscillations.
Our results establish a tractable platform for exploring interaction-induced temporal boundaries and extend the concepts of temporal reflection and refraction into a genuinely nonlinear regime.

\section{Model}
\begin{figure}[t]
    \centering
    \includegraphics[width=0.45\textwidth]{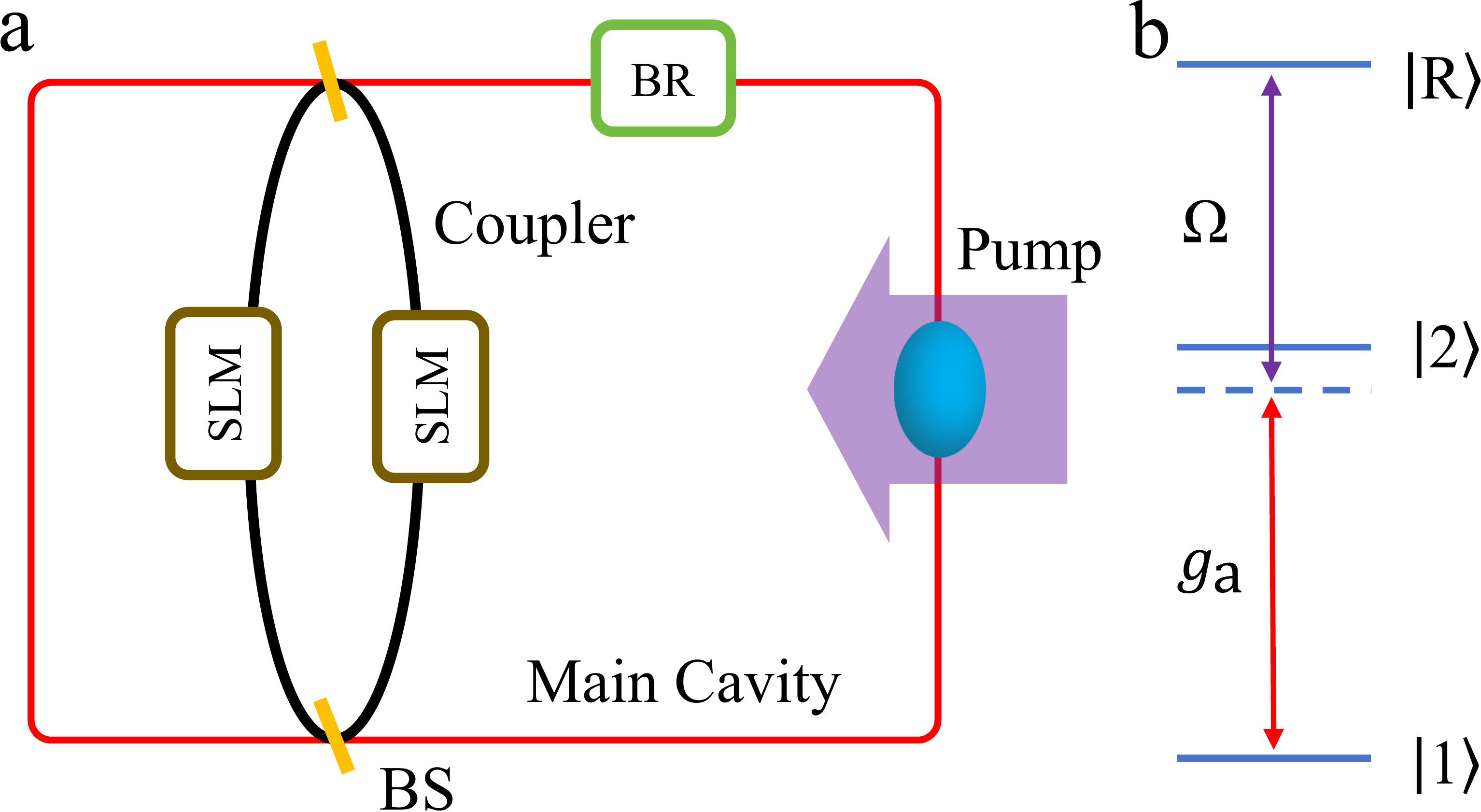}
    \caption{\justifying{\textbf{Illustration of the experimental design.} (a) Spatial light modulators (SLMs) inside the coupler cavities facilitate the coupling of OAM modes in the main cavity. (b) Nonlinearity emerges from the Raman-transition-based interaction between cavity photons and Rydberg atoms, driven by an external pump $\Omega$ alongside the atom-cavity coupling $g_a$.}}\label{fig:setups}
\end{figure}

Our proposal builds on the orbital angular momentum (OAM) synthetic-dimension platform demonstrated in Fig.~\ref{fig:setups}a, where the orbital angular momentum modes of a degenerate optical cavity serve as lattice sites~\cite{Luo2015QuantumSimulationa}. The azimuthal angle $\theta$ plays the role of the conjugate Bloch momentum, with the OAM mode number $l$ labeling the synthetic lattice coordinate.
Auxiliary cavities generate controllable nearest-neighbor tunneling at a rate $J$. This configuration establishes a synthetic single particle tight-binding model governed by the Hamiltonian~\cite{Luo2018TopologicalPhotonic}:
\begin{equation}
    \mathcal{H}_0=\sum_l J\hat{c}_{l+1}^\dagger \hat{c}_l + \text{h.c.},\label{eq:H0}
\end{equation}
where $J$ is the tunneling rate between each site with $\hat{c}_l^\dagger$ being the bosonic particle creation operator for synthetic site $l$.
The azimuthal angle $\theta$ and synthetic lattice site index $l$ are a pair of conjugate parameters. Therefore, here $\theta$ represent the synthetic Bloch momentum of the synthetic lattice. By Fourier transform $\hat c_\theta=\sum~\hat c_l e^{il\theta}/\sqrt{2\pi}$, the field operators are mapped from synthetic-lattice space $\hat{c}_l$ to the synthetic Bloch momentum space $\hat{c}_{\theta}$.

To dynamically engineer the nonlinearity, we rely on a Raman progress that couples the degenerate cavity photons to the highly excited Rydberg states of an atomic ensemble (see Fig.~\ref{fig:setups}b). By hybridizing the optical modes with these atomic excitations, we create cavity-Rydberg polaritons; this mechanism ultimately induces a state-dependent photon-photon interaction localized within the continuous $\theta$ space~\cite{vsibalic2018rydberg,Georgakopoulos2018TheoryInteractinga,chen2026topological}
\begin{equation}
    \mathcal{H}_\text{int}=g\int d\theta~\hat c_\theta^\dagger c_\theta^\dagger c_\theta c_\theta.
    \label{eq:Hint}
\end{equation}
Here $g$ is the interaction strength given explicitly by~\cite{chen2026topological}
\begin{equation}
    g\sim\frac{C_6}{R_B^5}\sin^4\left[\arctan\frac{g_a}{\Omega}\right],
\end{equation}
where $C_6$ is a coefficient depends on the Rydberg level we selected and $R_B$ is the blockade radius~\cite{Pritchard2010CooperativeAtomlight}, while $g_a$ and $\Omega$ are the collective coupling strength and Rabi frequency of the control field. Consequently, by simply tuning the frequency and intensity of the pump field, we can precisely control this unique interaction strength over a wide range, and even reverse its sign.
Unlike atom-momentum-based synthetic lattices~\cite{Wang2023TestingUniversality,An2018CorrelatedDynamics,Xie2020TopologicalQuantum,Chen2025InteractiondrivenBreakdown}, the interactions in our synthetic-lattice system act like all-to-all nonlocal interactions and preserve the total OAM~\cite{chen2026topological}.

Moreover, the effective gradient potential responsible for the Bloch dynamics can be implemented by inserting a beam rotator into the main cavity, which induces a controllable linear phase shift along the synthetic lattice dimension~\cite{Zhu2022SimulatingElectrical,Luo2015QuantumSimulation,Courtial1998RotationalFrequency},
leading to an effective Hamiltonian 
\begin{equation}
    \mathcal{H}_\text{B}=\sum_l\Lambda l\hat{c}_{l}^\dagger \hat{c}_{l}.
    \label{eq:Hg}
\end{equation}
The total Hamiltonian $\mathcal{H}_{\text{tot}}=\mathcal{H}_\text{0}+\mathcal{H}_\text{int}+\mathcal{H}_\text{B}$ in the Bloch momentum space is
\begin{equation}
    \mathcal{H}_\text{tot}=\int d\theta~2J\cos(\theta)\hat c^\dagger_\theta \hat c_\theta+g\hat c^\dagger_\theta \hat c^\dagger_\theta \hat c_\theta\hat c_\theta+\Lambda \frac{\partial}{i\partial \theta}\hat{c}_{\theta}^\dagger \hat{c}_{\theta}.
    \label{eq:Hint}
\end{equation}
The Heisenberg equation of motion for $\hat{c}_\theta$ can be derived as
\begin{equation}
    i\partial_t \hat c_\theta=2J\cos(\theta)\hat c_\theta+2g\hat c^\dagger_\theta \hat c_\theta\hat c_\theta+\Lambda \frac{\partial}{i\partial \theta}\hat{c}_{\theta}.
\end{equation}
The mean-field solution can be obtained by treating the field operators as c-numbers~\cite{Griffin1996BoseeinsteinCondensation,Dalfovo1999TheoryBoseeinstein,Smirnova2020NonlinearTopological,Stolen1973OpticalKerr,Gross1961StructureQuantized}, which represent the coherent-state approximation, and solving the nonlinear Schr$\ddot{\text{o}}$dinger equation
\begin{equation}
    i\partial_t \psi(\theta,t)=
    \left[2J\cos(\theta)+2U|\psi(\theta,t)|^2+\Lambda \frac{\partial}{i\partial \theta}\right]\psi(\theta,t),\label{eq:eom}
\end{equation}
where we normalize the wave function by letting $\psi(\theta,t)=\langle c_\theta\rangle/\sqrt{\rho_0}$ and $U=g\rho_0$.
The nonlinear dynamics governed by Eq.~\eqref{eq:eom} possess an unusual and highly constrained structure: the interaction term is local in Bloch momentum $\theta$ space.
Unlike conventional nonlinear lattice models, where interactions unavoidably induce mode mixing between different $\theta$ components, the present dynamics preserve the $\theta$-resolved density independently at each $\theta$. As a consequence, the nonlinearity generates only a $\theta$-dependent self-phase modulation, allowing the nonlinear evolution to remain exactly solvable.

\section{Dynamics and effective band}
To derive the analytical solution, we first move into the accelerated frame following the semiclassical Bloch trajectory by introducing $q=\theta-\Lambda t$, yielding
\begin{equation}
    i\partial_t \psi(q,t)=\left[2J\cos(q+\Lambda t)+2U|\psi(q,t)|^2\right]\psi(q,t).
\end{equation}
It is worth noticing that the Bloch momentum-resolved density remains conserved during the evolution,
\begin{equation}
    \partial_t |\psi(q,t)|^2=0.
\end{equation}
The nonlinear interaction therefore acts only as a Bloch momentum-dependent phase shift, allowing the equation of motion to be integrated analytically.
After transforming back to the laboratory frame, we obtain the exact solution (see Appendix~\ref{app:solution} for more details)
\begin{equation}
    \psi(\theta,t)=f(\theta-\Lambda t)e^{-i\phi_{\text{lin}}}e^{-i\phi_{\text{nl}}},\label{eq:solution}
\end{equation}
where 
\begin{eqnarray}
    \phi_{\text{lin}}&=&\frac{2J}{\Lambda}[\sin(\theta)-\sin(\theta-\Lambda t)], \\
    \phi_{\text{nl}}&=&2U|f(\theta-\Lambda t)|^2t,
\end{eqnarray}
are the linear and 
nonlinear phases, respectively. The initial wave packet is given by $\psi(\theta,0)=f(\theta)$, with normalization $\int d\theta |f(\theta)|^2=1$.
The wave packet in the synthetic lattice space can be obtained through Fourier transformation 
\begin{eqnarray}
    \psi(l,t)=\int d\theta~\psi(\theta,t) \frac{e^{-il\theta}}{\sqrt{2\pi}} 
\end{eqnarray}

In the absence of the gradient field, the nonlinear dynamics reduce to a static self-induced temporal medium characterized by an emergent wave-packet-dependent band structure.
This regime naturally gives rise to nonlinear temporal scattering phenomena including state-dependent refraction and temporal mode splitting.
However, without external confinement, the wave packet inevitably disperses during propagation, thereby limiting coherent control over long evolution times.
This limitation can be overcome by introducing a finite $\Lambda$.
In the following, we first investigate the limit $\Lambda\to 0$ to establish the fundamental nonlinear temporal boundary physics, and subsequently return to the finite-$\Lambda$ regime where Bloch dynamics suppress the dispersion and stabilize the nonlinear coherent temporal transport.

\begin{figure}[t]
    \centering
    \includegraphics[width=0.48\textwidth]{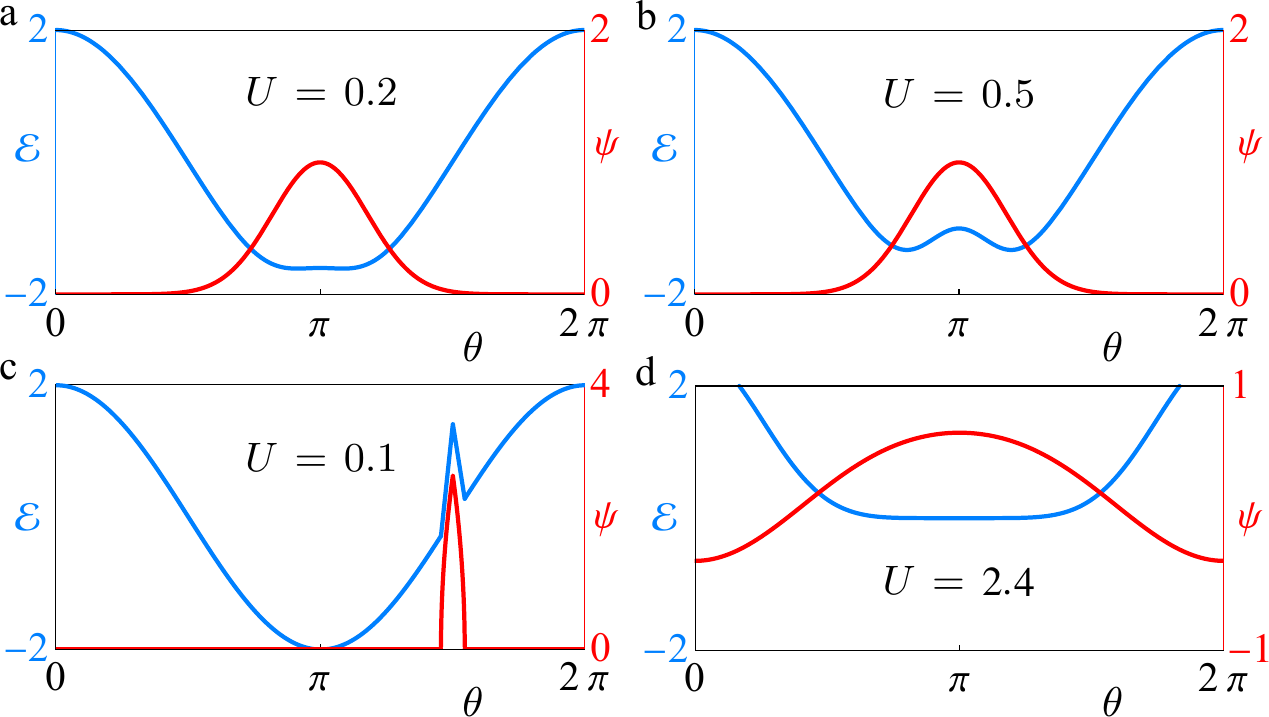}
	\caption{\justifying{\textbf{Effective nonlinear energy bands.} (a-b) The effective band of the Gaussian wave packet, $f(\theta)=(1/1.28\pi)^{1/4}\operatorname{exp}{\{-[(\theta-\pi)/0.8]^2\}}$, under different nonlinear interaction strengths $U$. As $U$ increases, the effective dispersion evolves from an almost flat band at $U=0.2$ to a strongly distorted nonparabolic band at $U=0.5$. (c) Effective band dispersion of a triangle-shaped wave packet, $f(\theta)=(\operatorname{max}\{\sqrt{50}- 50|\theta-3\pi/2|,0\})^{1/2}$, which gives rise to two distinct group velocities. (d) Effective band dispersion of the specially designed wave packet, $f(\theta)=[5-12\cos(\theta)-\cos(2\theta)]/\sqrt{195\pi}$, where the effective band becomes nearly flat over a broad range of $\theta$. We set $J=1$ for all panels.}}\label{fig:effectiveband}
\end{figure}

In the absence of the gradient field, the dynamics are governed entirely by the emergent nonlinear dispersion generated by the propagating wave packet itself. The linear phase reduces to
\begin{equation}
    \lim_{\Lambda\to0}\frac{\sin(\theta)-\sin(\theta-\Lambda t)}{\Lambda}=t\cos(\theta),
\end{equation}
such that the exact solution becomes
\begin{equation}
    \psi(\theta,t)=f(\theta)e^{-i\left[2J\cos(\theta)+2U|f(\theta)|^2\right]t}.
\end{equation}
The nonlinear evolution also preserves the Bloch momentum-space density as mentioned above, such that each Bloch momentum component acquires only a deterministic phase accumulation during the evolution.
The exact solution allows the nonlinear dynamics to be interpreted through an emergent effective dispersion relation,
\begin{equation}
    \mathcal{E}(\theta)=2J\cos(\theta)+2U|f(\theta)|^2.
\end{equation}
The emergent effective nonlinear band depends explicitly on both the linear band and intensity distribution of the propagating wave packet, while the interaction strength controlling the degree of band reconstruction, in contrast to conventional band structures determined solely by the underlying single-particle lattice Hamiltonian.
As shown in Figs.~\ref{fig:effectiveband}a and ~\ref{fig:effectiveband}b, a Gaussian profile wave-packet modify the band minima at different interaction strength. Moreover, a triangle-shaped wave-packet can change the band slope significantly, while a carefully designed cosine-like wave-packet may lead to a broad flat band regime, as shown in Figs.~\ref{fig:effectiveband}c and \ref{fig:effectiveband}d, respectively.

The corresponding group velocity is given by
\begin{gather}
    v_g(\theta)=\partial_\theta \mathcal{E}(\theta),
\end{gather}
which now acquires an explicit dependence on the wave-packet profile and the nonlinear strength. Therefore, wave packets with different amplitudes or $\theta$-space profiles propagate along distinct trajectories even within the same physical medium.
This interaction-induced dispersion and group velocity naturally define a self-induced nonlinear temporal medium.
In contrast to conventional temporal boundaries where the refractive properties are externally prescribed material parameters, the effective temporal response here is dynamically generated by the propagating wave itself and the nonlinear strength, motivating an effective state-dependent temporal refractive index (see below) and establishing a nonlinear temporal scattering regime fundamentally distinct from conventional passive temporal media.

\section{Temporal boundary}
The wave-packet-dependent band structure reshapes temporal scattering at nonlinear boundaries.
Since the effective group velocity and refractive response are self-induced, abrupt temporal modulation of the nonlinear parameters no longer generates conventional passive temporal boundary effects, which are governed by externally prescribed material parameters, but instead leads to intrinsically nonlinear and state-dependent temporal boundary dynamics.
We consider a nonlinear temporal boundary generated by a sudden quench of the interaction strength $U_0\to U_1$, which reconstructs the emergent band structure from $\mathcal{E}_0(\theta)\to \mathcal{E}_1(\theta)$.
The $\theta$ distribution remains continuous across the temporal boundary, while the effective group velocity undergoes a discontinuous reconstruction $v_{g,0}(\theta)\to v_{g,1}(\theta)$ due to the modified nonlinear dispersion. 
The subsequent propagation is therefore determined by the post-quench group velocity.
Following the standard temporal-scattering picture, we classify the post-quench propagation as temporal refraction or reflection depending on whether the group velocity preserves or reverses its propagation direction across the boundary.

\begin{figure}[t]
    \centering
    \includegraphics[width=0.48\textwidth]{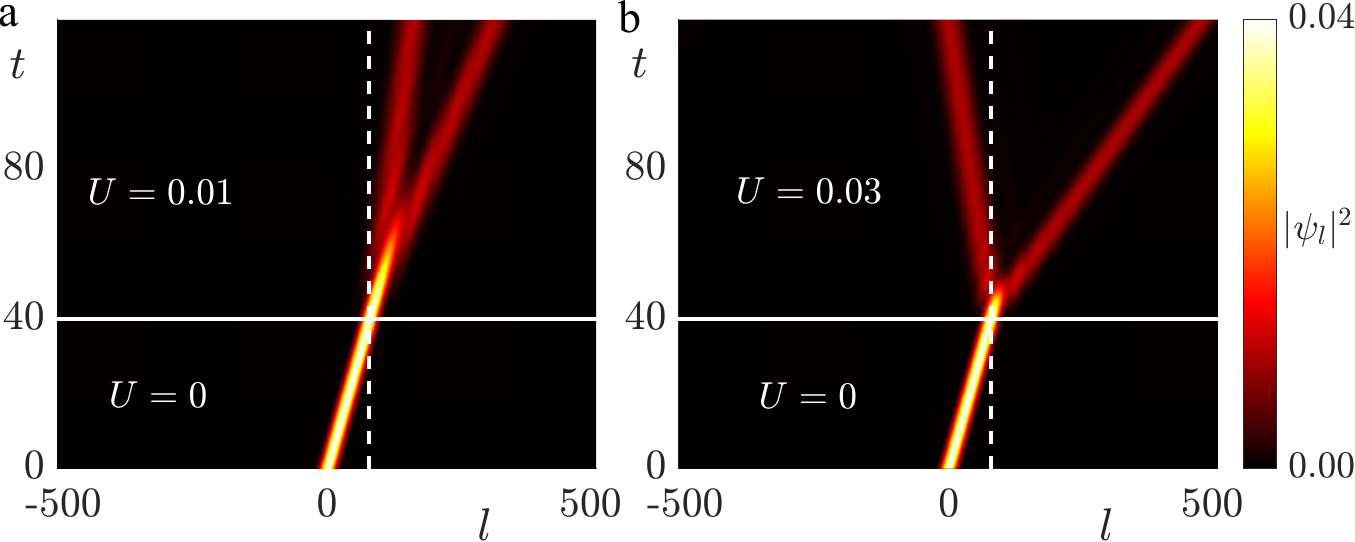}
	\caption{\justifying{\textbf{Temporal reflection and refraction.} The spacetime evolution of the wave packets is shown before and after the temporal boundary (white solid lines) at $t=t_B=40$. (a) For weak interaction ($U=0.01$), both trajectories after the boundary are on the same side (birefringence-like trajectory). (b) As the interaction strength increases ($U=0.03$), the scattered waves split to opposite sides. Parameters are fixed at $\alpha=50$, $J=1$ and $\Lambda=0$. The nonlinear strength equals to 0 before the temporal boundary.}}\label{fig:timebound}
\end{figure}

Furthermore, for a wave packet which exhibit distinct local structures at different Bloch momentum regions, the effective dispersion may support multiple propagation branches with distinct group velocities within a finite Bloch momentum window. Consequently, the wave packet splits into multiple temporal transport channels after crossing the boundary while these channels are modulated by the nonlinear strength.
Fig.~\ref{fig:timebound} presents a schematic spacetime evolution of an incident wave packet encountering a temporal boundary, the initial profile of the normalized triangle wave packet is
$f(\theta)=\left(\operatorname{max}\{\alpha^{1/2}-\alpha\left|\theta-3\pi/2\right|,0\}\right)^{1/2}$ with the slope $\alpha$.
At the moment of the abrupt parameter modulation ($t=t_B$), the incident wave packet splits into two independent components.
In Fig.~\ref{fig:timebound}a, the reconstructed band structure differs only slightly from the original dispersion for weak nonlinearities, leading to multiple forward-propagating scattering branches with distinct velocities, analogous to birefringence.
As the interaction strength increases, the nonlinear band becomes strongly distorted and the reconstructed group velocity may reverse sign, producing temporal reflection in addition to refraction as shown in Fig.~\ref{fig:timebound}b.

Using the effective group velocity, we define an effective temporal refractive index following the conventional temporal-optics analogy where the refractive index is inversely related to the propagation velocity which recovers the conventional temporal refraction relation derived for passive media~\cite{Dong2024QuantumTime} as
\begin{equation}
    n_{\text{eff},i}(\theta)v_{g,i}(\theta)=n_0v_0,
\end{equation}
where $n_0v_0$ is an arbitrarily chosen normalization constant. The corresponding temporal refraction relation then still takes the form like
\begin{equation}
    \frac{n_{\text{eff},1}}{n_{\text{eff,0}}}=\frac{\cos\beta_1}{\cos\beta_0},
\end{equation}
where $n_0$ ($n_1$) is the effective refractive index for $t<t_B$ ($t>t_B$) and $\beta_0$, $\beta_1$ are the incidence and refraction angles.
\begin{figure}[t]
    \centering
    \includegraphics[width=0.48\textwidth]{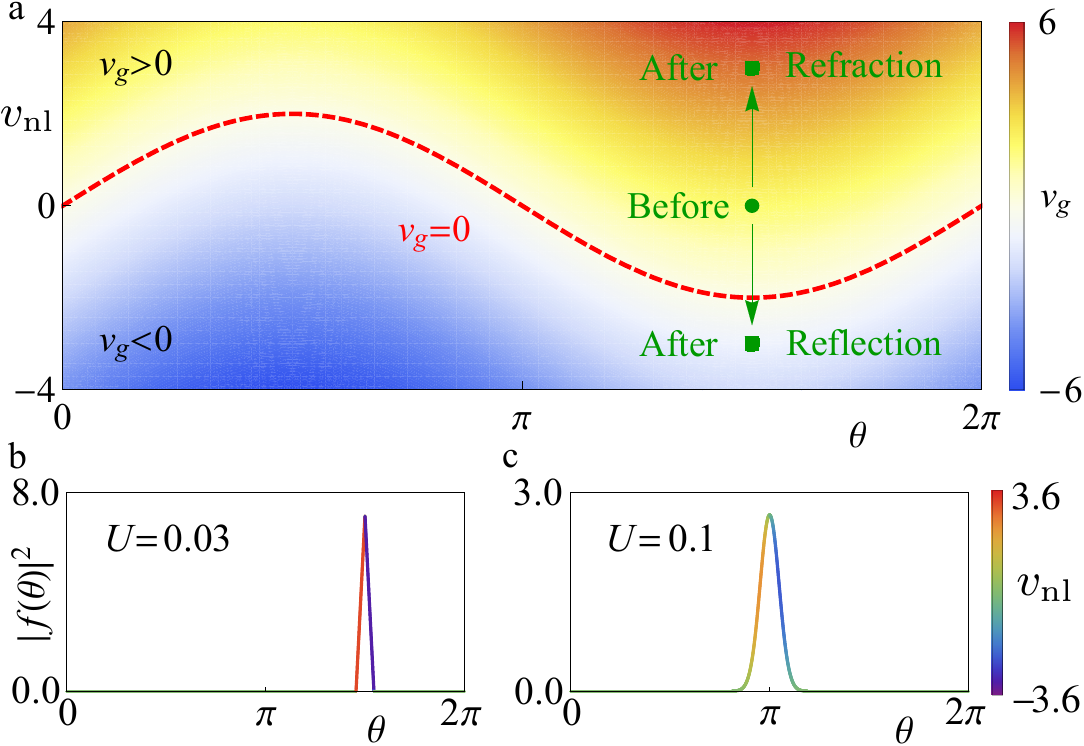}
	\caption{\justifying{\textbf{Scattering phase diagram.} (a) Group velocity in the $v_{\rm nl}$-$\theta$ plane. Refraction and reflection are determined by whether the group velocity changes its sign. The red dashed line corresponds to the boundary where $v_g$ changes its sign, leading to opposite propagation directions on the two sides. Temporal boundary splits the wave packet (indicated by the green dot) into two branches (i.e., the reflection and refraction indicated by two green squares), corresponding to the scattering process in Fig.~\ref{fig:timebound}b. Color bar indicates the total group velocity $v_g$. (b) and (c) Wave packet and the corresponding $v_\text{nl}$ for $f(\theta)=\text{max}\{(\sqrt{50}- 50|\theta-3\pi/2|)^{1/2},0\}$ in (b) and $f(\theta)=(\pi/0.045)^{1/4}\operatorname{exp}{\{-[(\theta-\pi)/0.09]^2\}}$ in (c). Color bar indicates the nonlinear group velocity $v_\text{nl}$. Here we set $J=1$ for all pannels.}}\label{fig:phase}
\end{figure}
Here, we divided the total group velocity into two parts to represent the contribution of the linear/nonlinear part of the effective dispersion relation.
\begin{equation}
    v_g=v_\text{l}+v_\text{nl}=-2J\sin\theta+2U\partial_\theta|f(\theta)|^2.
\end{equation}
The linear part remains unchanged because it is fixed by the single-particle Hamiltonian. In contrast, the nonlinear part depends on both the wavepacket profile and the strength of the nonlinearity, so it can be controlled.
Different scattering channels therefore occupy distinct regions in the nonlinear parameter space, giving rise to a temporal scattering phase diagram as shown in Fig.~\ref{fig:phase}a.
The green dot and squares represent the states before and after encountering the temporal interface in Fig.~\ref{fig:timebound}b, respectively.
After the temporal interface, the wave splits into components with opposite directions and different group velocities.
In the case of a triangular initial wavefunction, only two dominant slope values appear as shown in Fig.~\ref{fig:phase}b, whereas for a Gaussian wavepacket there exists a continuous and broadened slope distribution as shown in Fig.~\ref{fig:phase}c. This indicates that, 
the propagation depends not only on the nonlinear strength but also on the wave-packet profile in $\theta$-sapce.

The transition between temporal refraction and reflection is entirely determined by the sign change of the effective group velocity. It is important that the critical boundary with vanishing group velocity is therefore analytically given by
\begin{equation}
    v_\text{nl}=2J\sin(\theta).
\end{equation}
Note that even in the most general case with a initial non-zero nonlinear strength, the behavior of the wave can be obtained by calculating $v_g$ before and after the temporal boundary. Whether light waves are refracted or reflected depends on whether, across the temporal boundary, the group velocity changes its sign by crossing the critical boundary.

\section{Nonlinear Bloch transport}
Although the self-induced temporal medium enables highly tunable nonlinear scattering, long-time coherent transport in the static regime is ultimately limited by wave-packet spreading.
Since the effective nonlinear dispersion generally possesses finite curvature, long-time evolution inevitably leads to delocalization and degradation of the temporal scattering signal.
This problem forces us, even when conducting single particle experiments on refraction and reflection at time boundaries, to use wave packets that are extremely narrow in Bloch momentum space, 
which impose severe experimental constraints and compromise measurement fidelity.
A natural question therefore arises: can nonlinear temporal scattering remain coherent under long-time transport?
To answer this problem, we now return to the finite-gradient regime, where the accelerated Bloch dynamics fundamentally reshape the nonlinear transport.

Under the constant gradient field, the Bloch momentum follows the accelerated trajectory $\theta(t)=\theta(0)+\Lambda t$ which periodically traverses the Brillouin zone with Bloch period $T_B=2\pi/\Lambda$.
Importantly, the role of the original lattice dispersion changes in the finite-gradient regime. In conventional single particle dynamics, the bare band structure directly determines the group velocity and transport properties.
Recalling the exact solution \eqref{eq:solution}, however, the accelerated motion induced by the gradient field converts the lattice dispersion into a periodic contribution $\phi_\text{lin}$, which is a stabilizing mechanism enforcing periodic $\theta$ space cycling.
Consequently, the single-particle lattice band no longer directly governs the net transport dynamics. Instead, the gradient field enforces periodic propagation, while the actual temporal scattering behavior emerges entirely from the self-induced nonlinear dispersion.

Such dynamics allow precise engineering of the wave-packet trajectory switching between reflection and refraction modes, while ensuring that the quantum state is exactly restored at the end of each cycle without any time-reversal protocols or dynamical freezing, providing a robust platform for the potential applications discussed below.
\begin{figure}[t]
    \centering
    \includegraphics[width=0.48\textwidth]{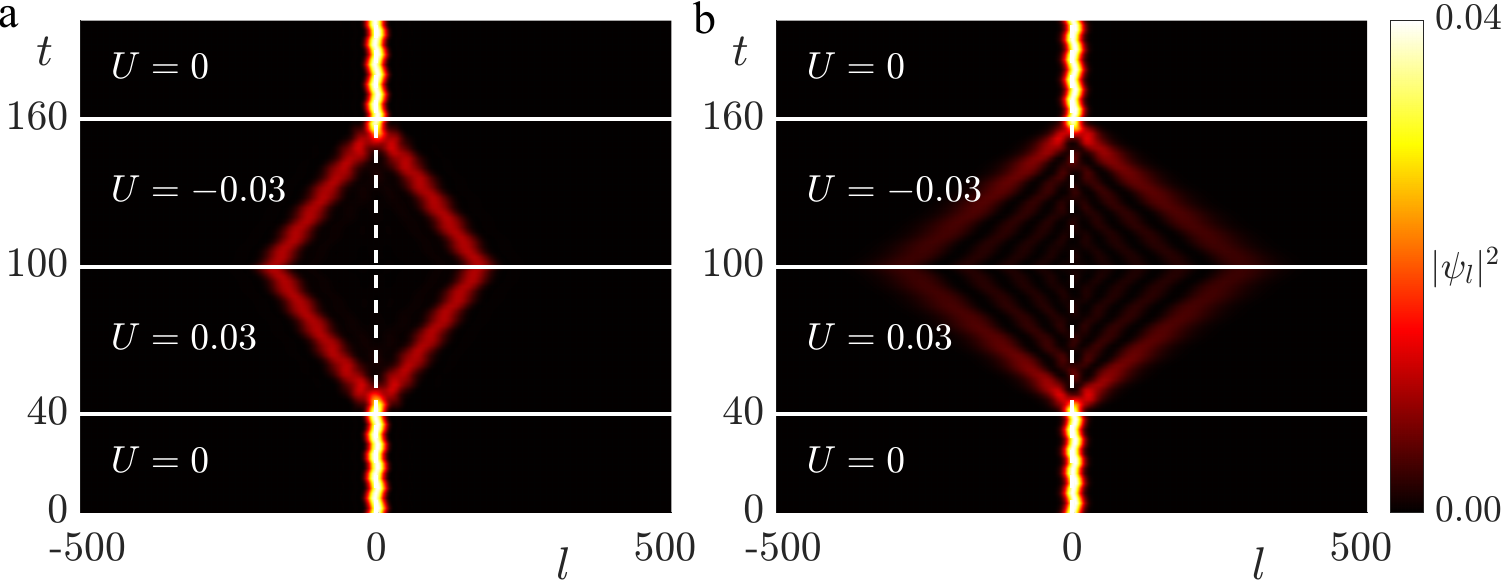}
	\caption{\justifying{\textbf{Perfect temporary restoration enabled by Bloch dynamics.} Spacetime evolution of (a) a triangular $f(\theta)=\operatorname{max}\{\left(\sqrt{50}-50\left|\theta-\pi/2\right|\right)^{1/2},0\}$ and (b) a Gaussian $f(\theta)=(0.005\pi)^{1/4}e^{-\theta^2/0.01}$ initial state. By changing the effect nonlinear strength twice at $t=40~\text{and}~100$, both distinct wave packets undergo exact reconstruction at the end of the whole cycle ($t=160$), demonstrating the robust, shape-independent perfect restoration of the system. Common parameters $J=1$ and $\Lambda=0.75$.}}\label{fig:bloch}
\end{figure}
One special combination of the temporal boundaries with this perfect recovery of the quantum state bears qualitative similarities to diffraction-free temporal imaging. A triangle-shaped wave package as shown in Fig.~\ref{fig:bloch}a only has two main slope values, leading to a split and restore, while a Gaussian wave package in Fig.~\ref{fig:bloch}b has a rich slope distribution and cause a complex interference during the propagation. Both of them encounter three temporal boundaries for the recovery.
Moreover, the effective interaction strength $U=g\rho_0$ is associated with the density of the wave function.
Consequently, when subjected to a temporal boundary, signals of different intensities exhibit divergent scattering behaviors.
For instance, a low-intensity signal (small $\rho_0$) predominantly governed by the linear hopping term $J$ may undergo standard time-refraction, continuing along its original direction with minimal deviation.
In contrast, a high-intensity pulse (large $\rho_0$) experiences a strongly distorted effective band, leading to pronounced temporal birefringence or even complete time-reflection.
The coexistence of exact nonlinear transport and tunable temporal scattering makes the present system particularly promising for synthetic photonic implementations.

\section{Experimental feasibility}
Assuming a standard cavity length of roughly $0.3$ m, the free spectral range $\Omega_{\rm FSR}$ falls near $10$ GHz. By selecting a beam splitter with a reflectivity $r \approx 0.1$, the synthetic hopping rate $J$ scales to the $10$ MHz regime since the tunneling rate $J \sim r \Omega_{\rm FSR}/4\pi$~\cite{Luo2018TopologicalPhotonic}. While typical Rydberg interactions reach the MHz scale at micrometer separations~\cite{vsibalic2018rydberg}, the system is well-equipped to support strong nonlinearity as $U\sim 10^{-1} J$ with increasing photon number~\cite{Labuhn2016TunableTwodimensional,Bernien2017ProbingManybody}.

\begin{figure}[t]
    \centering
    \includegraphics[width=0.48\textwidth]{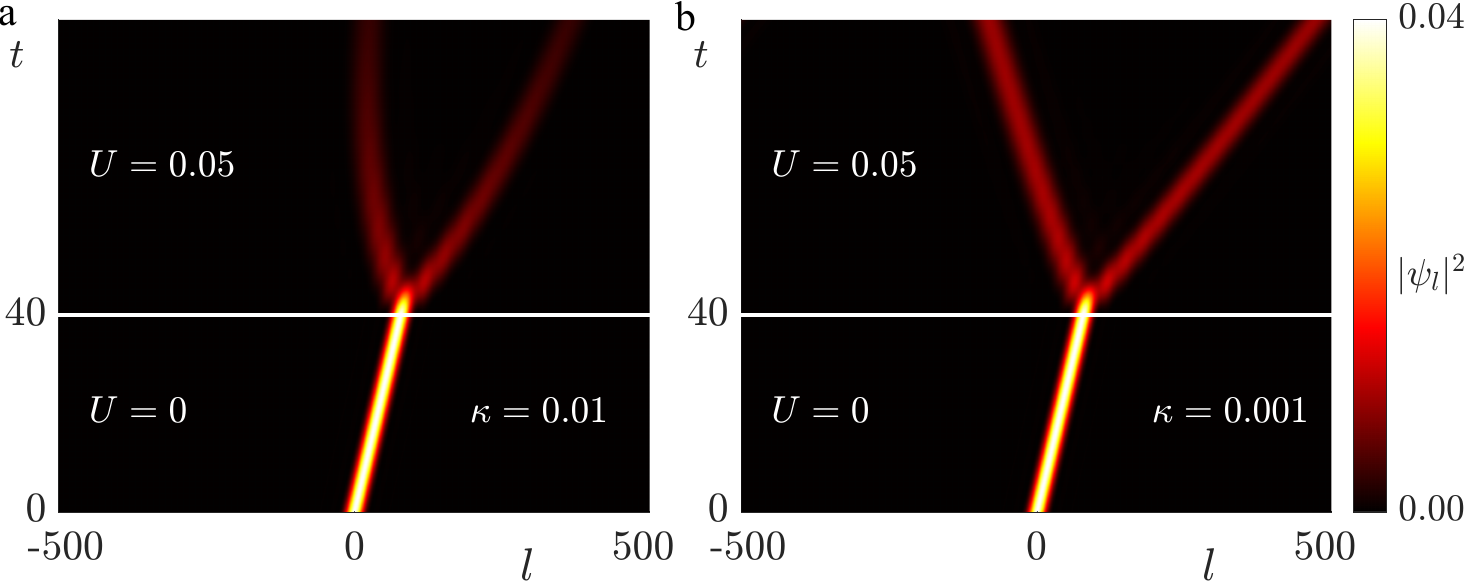}
    \caption{\justifying{\textbf{Temporal boundary scattering under different photon losses.} (a-b) Evolution of a wave packet with a nonlinear of $U=g\rho_0=0.05J$ quench at the temporal boundary with different decay rates $\kappa$. For $\kappa=0.01J$, the dynamics persists and there is still clear refraction and reflection phenomenon during the brief period immediately following the encounter with the boundary. We set $J=1$ and the initial profile $f(\theta)=\left(\operatorname{max}\{50^{1/2}-50\left|\theta-3\pi/2\right|,0\}\right)^{1/2}$.}}\label{fig:decay}
\end{figure}

The fundamental limit to observing coherent temporal scattering is the photon lifetime governed by the cavity linewidth $\kappa_c$. 
Because the temporal scattering phenomena rely critically on the density-dependent effective nonlinearity, photon leakage from the cavity progressively weakens the nonlinear response and can ultimately change the temporal scattering.
To ensure that the nonlinear temporal scattering phenomenon survives before cavity photons decay significantly, the duration of the entire scattering event at the temporal boundary must remain well within the photon lifetime, a condition that sets a lower bound on the required gradient strength. Here we find a realistic linewidth $\kappa\simeq 0.01 J\simeq 0.1$ MHz is small enough to ensure the birefringence happens, with an effective nonlinearity quench ($U=0.05J$) at the temporal boundary.
In the case of a large decay rate, the wavefunction still exhibits a refracted component immediately after encountering the temporal interface. However, as time increases, the reduction of the effective nonlinear strength induced by the decay causes the wavefunction to gradually recover its initial propagation direction as shown in Fig.~\ref{fig:decay}a.
Meanwhile, the resulting change in group velocity becomes negligible as shown in Fig.~\ref{fig:decay}b if the linewidth can be improved to $\kappa\lesssim0.001J\sim$ 10kHz by using high-performance optical elements~\cite{PhysRevLett.121.220405}. It is worth noting that the loss rate may change the dynamics cross over to another situation during the propagation.

Although our analysis has focused on interactions that are local in Bloch momentum space (equivalently, contact interactions in the $\theta$ representation), the framework can be generalized straightforwardly to finite-range interactions in $\theta$ space, while preserving the essential features of nonlinear temporal scattering. In the presence of a nonlocal interaction, the interaction Hamiltonian takes the form
\begin{equation}
    H_{\rm int}=g\int d\theta d\theta'~c^\dagger_\theta c^\dagger_{\theta'}F(|\theta-\theta'|) c_{\theta'} c_{\theta},\label{eq:gHint}
\end{equation}
where $F(|\theta-\theta'|)$ is a smooth interaction kernel that depends only on the relative distance. In this case, the exact analytical solution retains the same structure as in the local-interaction limit, with the nonlinear phase accumulation generalized to
\begin{eqnarray}
    \phi_{\text{nl}}=2Uf_2(\theta-\Lambda t)t
\end{eqnarray}
with $f_2(\theta)=\int d\theta' F(|\theta-\theta'|)|f(\theta')|^2$ (see Appendix~\ref{app:solution} for more details). The nonlinear dynamics can still be interpreted in terms of an emergent effective band structure because the momentum-resolved density remains conserved throughout the evolution,
\begin{equation}
    \partial_t |\psi(\theta-\Lambda t,t)|^2=0.
\end{equation}
As a consequence, the effective reconstruction of the group velocity and the associated temporal reflection and refraction phenomena remain qualitatively unchanged. Finite-range interactions modify only the quantitative form of the effective group velocity $v_g(\theta)$ and its dependence on the wave-packet profile. Note that, $F$ becomes a delta function for a contact interaction, and $f_2(\theta)$ reduces to $|f(\theta)|^2$. More generally, the exact solvability of the present model does not rely on strictly local momentum-space interactions, but rather on the conservation of the momentum-resolved density, which guarantees a self-consistent description of the nonlinear dynamics in terms of an emergent band structure.

\section{Conclusion and discussion}
In conclusion, we have investigated nonlinear temporal boundaries generated by interaction quenches in synthetic lattices and developed an exactly solvable framework for nonlinear temporal scattering. We obtained analytical solutions for the evolution of arbitrary wave packets and showed that the nonlinear dynamics can be understood in terms of an emergent effective band structure whose dispersion, group velocity, and refractive properties are self-consistently determined by the wave packet itself. This self-induced temporal medium gives rise to intrinsically state-dependent scattering phenomena, including amplitude-dependent temporal reflection/refraction and nonlinear temporal birefringence.
We further demonstrated that gradient-induced Bloch oscillations suppress wave-packet diffusion and enable coherent transport with exact state reconstruction over long evolution times. The coexistence of controllable nonlinear temporal scattering and robust coherent transport provides a versatile route toward engineered wave manipulation at interaction-induced temporal boundary.

Beyond the specific interaction considered here, we showed that the exact solvability of the model extends naturally to a broader class of finite-range interactions. The underlying principle is the conservation of the momentum-resolved density, which allows the nonlinear evolution to be described through a self-consistent emergent band structure. Our work therefore extends the concepts of temporal reflection and refraction from dispersion-quenched linear systems to interaction-quenched nonlinear media, and establishes a promising framework for exploring nonlinear temporal scattering, nonequilibrium transport, and interaction-driven wave dynamics.

\section*{Acknowledgment}
We thank Han Pu for insightful discussions. This work is supported by the National Natural Science Foundation of China (Grants No. 12574544 and No. 12474366) and Quantum Science and Technology-National Science and Technology Major Project (Grant No. 2021ZD0301200).
X.-W.Luo also acknowledges the support from USTC start-up funding.

\bibliography{maincite}

\appendix

\section{Exact Solution of Equation of Motion}\label{app:solution}
We derive the exact solution from the single-particle Hamiltonian Eq.~\eqref{eq:H0} with a gradient potential Eq.~\eqref{eq:Hg} and the general interaction Eq.~\eqref{eq:gHint} starting from the total Hamiltonian
\begin{gather}
    H_\text{tot}=\int d\theta~2J\cos(\theta)\hat c^\dagger_\theta \hat c_\theta+\Lambda \frac{\partial}{i\partial \theta}\hat{c}_{\theta}^\dagger \hat{c}_{\theta}\nonumber\\
    +g\int d\theta d\theta'~c^\dagger_\theta c^\dagger_{\theta'}F(|\theta-\theta'|) c_{\theta'} c_{\theta}.
\end{gather}
The dynamics of the field operators can be obtained through the Heisenberg equation
\begin{gather}
    i\frac{\partial}{\partial t}c_\theta=[c_\theta^\dagger,H_{\rm tot}].
\end{gather}
By replacing the operators with the mean-field c-numbers, we can obtain the nonlinear Schr$\ddot{\text{o}}$dinger equation as
\begin{eqnarray}
i\partial_t &&\psi(\theta,t)=\left[2J\cos\theta+\Lambda \frac{\partial}{i\partial \theta}\right.\nonumber\\
&&\left.+2U\int d\theta'~F(|\theta-\theta'|)|\psi(\theta',t)|^2\right]\psi(\theta,t),
\label{eq:A1}
\end{eqnarray}
with initial condition $\psi(\theta,0)=f(\theta)$.
To solve Eq.~(\ref{eq:A1}), we first introduce a comoving coordinate
\begin{equation}
q=\theta-\Lambda t,
\end{equation}
which removes the linear drift term by using the chain rule,
\begin{eqnarray}
\partial_t\psi(\theta,t)&=&\partial_t\psi(q,t)-\Lambda\partial_q\psi(q,t),
\\
\partial_\theta\psi(\theta,t)&=&\partial_q\psi(q,t).
\end{eqnarray}
Substituting into Eq.~(\ref{eq:A1}), we obtain
\begin{eqnarray}
i\partial_t &&\psi(q,t)=\left[2J\cos(q+\Lambda t)\right.\nonumber\\
&&\left.+2U\int dq'~F(|q-q'|)|\psi(q',t)|^2\right]\psi(q,t),
\label{eq:A2}
\end{eqnarray}
where the drift term $\Lambda \partial_\theta/i$ cancels exactly. Importantly, Eq.~(\ref{eq:A2}) implies a local conservation law for the density:
\begin{equation}
\partial_t |\psi(q,t)|^2=\psi^*\partial_t \psi+\psi \partial_t \psi^*=0,
\end{equation}
since the right-hand side of Eq.~(\ref{eq:A2}) is purely multiplicative and contributes only to the phase evolution of $\psi(q,t)$. Therefore, the envelope of the profile is strictly time-independent in the comoving frame, and no self-modulation of the wave-packet shape occurs.

Eq.~\eqref{eq:A2} is now local in $q$. It has no spatial derivatives, and therefore its solution can be written in amplitude-phase form
\begin{equation}
\psi(q,t)=f(q)\exp[-i\Phi(q,t)],
\end{equation}
with $f(q)=\psi(q,0)$ fixed by the initial condition. Substituting this ansatz into Eq.~(\ref{eq:A2}), we obtain a purely phase evolution equation
\begin{eqnarray}
\partial_t \Phi(q,t)&=&2J\cos(q+\Lambda t)\nonumber\\
&+&2U\int dq'~F(|q-q'|)|\psi(q',t)|^2.
\end{eqnarray}
Integrating over time gives
\begin{equation}
\Phi(q,t)=2Uf_2(q)t+\frac{2J}{\Lambda}\left[\sin(q+\Lambda t)-\sin(q)\right],
\end{equation}
where $f_2(q)=\int dq' F(|q-q'|)|f(q')|^2$ is the effective static nonlinear potential. Returning to the original coordinate $q=\theta-\Lambda t$, the rigorous analytical solution becomes
\begin{equation}
\psi(\theta,t)=f(\theta-\Lambda t)\exp\!\left[-i\phi_{\text{lin}}\right]\exp\!\left[-i\phi_{\text{nl}}\right],
\end{equation}
where
\begin{eqnarray}
\phi_{\text{lin}}&=&\frac{2J}{\Lambda}\left[\sin\theta - \sin(\theta-\Lambda t)\right],\\
\phi_{\text{nl}}&=&2Uf_2(\theta-\Lambda t)t.
\end{eqnarray}
Specifically, $F(|\theta-\theta'|)$ becomes a delta function for a local interaction in $\theta$-space (i.e., Eq.~\eqref{eq:Hint}). And the nonlinear phase becomes
\begin{eqnarray}
    \phi_{\text{nl}}&=&2U\int dq' \delta(\theta-\Lambda t-q')|f(q')|^2t\nonumber\\
    &=&2U|f(\theta-\Lambda t)|^2t,
\end{eqnarray}
which is the same as Eq.~(\ref{eq:solution}) in the main text.
\end{document}